\documentclass[aps,prl,twocolumn]{revtex4-1}
\usepackage{amsmath,amssymb,bm,epsfig,graphicx,longtable}
\begin{document}

\title{Density-gradient-free variable in exchange-correlation functionals
for detecting inhomogeneities in the electron density}
\author{Fabien Tran}
\author{Peter Blaha}
\affiliation{Institute of Materials Chemistry, Vienna University of Technology,
Getreidemarkt 9/165-TC, A-1060 Vienna, Austria}

\begin{abstract}

An alternative type of approximation for the exchange and correlation functional in density
functional theory is proposed. This approximation depends on a variable $u$
that is able to detect inhomogeneities in the electron density $\rho$ without
using derivatives of $\rho$. Instead, $u$ depends on the orbital
energies which can also be used to measure how a system differs from
the homogeneous electron gas. Starting from the functional of Perdew, Burke,
and Ernzerhof (PBE) [Phys. Rev. Lett. \textbf{77}, 3865 (1996)], a
functional depending on $u$ is constructed. Tests on the lattice constant,
bulk modulus, and cohesive energy of solids show that this $u$-dependent
PBE-like functional is on average as accurate as the original PBE or
its solid-state version PBEsol. Since $u$ carries more nonlocality than the
reduced density gradient $s$ used in functionals of the generalized gradient
approximation (GGA) like PBE and $\alpha$ used in meta-GGAs,
it will be certainly useful for the future development of more accurate
exchange-correlation functionals.

\end{abstract}

\maketitle

Kohn-Sham density functional theory (KS-DFT)
\cite{HohenbergPR64,KohnPR65} is the most used quantum mechanical method to
calculate the geometry and electronic structure of molecules, surfaces, and
solids \cite{CohenCR12,BeckeJCP14}. The success of KS-DFT is due to its ratio
cost/accuracy which is quite low compared to other methods, particularly
when the exchange-correlation (xc) functional $E_{\text{xc}}$ is of
semilocal type. This allows us to treat up to several thousands
of atoms routinely.
The semilocal functionals $E_{\text{xc}}$ belong
to the first three rungs of Jacob's ladder \cite{PerdewAIP01}
and the xc-energy density
$\varepsilon_{\text{xc}}$, defined as
\begin{equation}
E_{\text{xc}} = \int\varepsilon_{\text{xc}}(\mathbf{r})d^{3}r,
\label{Exc}
\end{equation}
depends locally on properties of the system.
In the local density approximation (LDA, first rung of
Jacob's ladder) \cite{KohnPR65},
$\varepsilon_{\text{xc}}$ is a function of the electron density
$\rho=\sum_{i=1}^{N}\left\vert\psi_{i}\right\vert^{2}$, while
in the generalized gradient approximation (GGA, second rung of Jacob's ladder)
\cite{BeckePRA88,PerdewPRL96},
$\varepsilon_{\text{xc}}$ depends on $\rho$ and its first derivative
$\nabla\rho$. In meta-GGA (third rung of Jacob's ladder)
\cite{VanVoorhisJCP98,PerdewPRL99,DellaSalaIJQC16},
$\varepsilon_{\text{xc}}$ depends additionally
on $\nabla^{2}\rho$ and/or the kinetic-energy density
$\tau=\left(1/2\right)\sum_{i=1}^{N}\nabla\psi_{i}^{*}\cdot\nabla\psi_{i}$.

It is clear that using more and more ingredients ($\rho$, $\nabla\rho$,
$\tau$, etc.) for the construction of an xc functional should increase
the overall accuracy, and studies have shown that it is the case (see, e.g.,
Refs.~\cite{MardirossianJCP15,DellaSalaIJQC16,TranJCP16} for recent works). 
For instance, the recent meta-GGA SCAN
(strongly constrained and appropriately normed) \cite{SunPRL15}
has shown to be quite broadly accurate for properties depending
on the total energy \cite{PengPRX16,ZhangNJP18}, although some
problems remain \cite{GarzaJCTC18,IsaacsPRM18}.

In this work, we propose an alternative type of approximation for
$\varepsilon_{\text{xc}}$ in Eq.~(\ref{Exc}), which depends on
\begin{equation}
u(\mathbf{r})=
B\sum_{i=1}^{N}\sqrt{\epsilon_{\text{H}}-\epsilon_{i}}
\frac{\left\vert\psi_{i}(\mathbf{r})\right\vert^{2}}{\rho^{4/3}(\mathbf{r})},
\label{u}
\end{equation}
where $\epsilon_{i}$ are the energies of the orbitals ($i=\text{H}$ is the
highest occupied one) and $B=16\sqrt{2}\pi/\left(3\pi^2\right)^{4/3}$.
The particularity of Eq.~(\ref{u}) is to be $\epsilon_{i}$ dependent and,
to our knowledge, the only existing xc-energy functionals
that depend on the orbital energies $\epsilon_{i}$ are those derived
from \textit{ab initio} methods, e.g., perturbation theory or
the random phase approximation \cite{EngelJCC99,Engel}.
Actually, $u$ can be expressed as
$u=v_{\text{x,resp}}^{\text{GLLB}}/v_{\text{x,resp}}^{\text{HEG}}$,
where $v_{\text{x,resp}}^{\text{GLLB}}$ is the approximation to
the response term of the exchange
potential proposed by Gritsenko \textit{et al}. (GLLB) \cite{GritsenkoPRA95}
and $v_{\text{x,resp}}^{\text{HEG}}=k_{\text{F}}/\left(2\pi\right)$,
where $k_{\text{F}}=\left(3\pi^{2}\rho\right)^{1/3}$,
is the exact homogeneous electron gas (HEG) limit of
$v_{\text{x,resp}}$. Since $v_{\text{x,resp}}^{\text{GLLB}}\rightarrow
v_{\text{x,resp}}^{\text{HEG}}$ in the HEG-limit (see Ref.~\cite{GritsenkoPRA95}),
$u=1$ for the HEG and
any departure from this value indicates that somewhere in the system the
density $\rho$ is not constant.

Two interesting features of Eq.~(\ref{u}) should be mentioned:
(a) It can detect inhomogeneities in $\rho$ without explicitly using
the derivatives of $\rho$ and (b) it does not make the calculation of
$\varepsilon_{\text{xc}}$ more expensive than for semilocal methods.
Thus, $u$ could be considered as an alternative or a complement
to functions which depend explicitly on derivatives of $\rho$, e.g.,
the reduced density gradient
\begin{equation}
s(\mathbf{r})=\frac{\left\vert\nabla\rho(\mathbf{r})\right\vert}
{2\rho(\mathbf{r})k_{\text{F}}(\mathbf{r})},
\label{s}
\end{equation}
in GGA functionals \cite{BeckeJCP86a,PerdewPRB86},
or $\tau$-dependent functions like
\begin{equation}
\alpha(\mathbf{r})=\frac{\tau(\mathbf{r})-\tau^{\text{W}}(\mathbf{r})}
{\tau^{\text{TF}}(\mathbf{r})},
\label{alpha}
\end{equation}
where
$\tau^{\text{W}}=\left\vert\nabla\rho\right\vert^{2}/\left(8\rho\right)$ and
$\tau^{\text{TF}}=\left(3/10\right)\left(3\pi^{2}\right)^{2/3}\rho^{5/3}$ are
the von Weizs\"{a}cker \cite{vonWeizsackerZP35} and Thomas-Fermi
kinetic-energy density \cite{ThomasPCPS27,FermiRANL27}, that is used
in meta-GGAs \cite{BeckeJCP90,TaoPRL03,SunPRL13}.
Note that, similarly as $\alpha$, $u=0$
for one- and spin-compensated two-electron systems, such that $u$
can also be used to eliminate the self-interaction error.

In the aim of showing the usefulness of $u$ as a variable in xc-energy
functionals and its potential interest for improving further the accuracy
of fast DFT methods, we construct an xc-energy density
$\varepsilon_{\text{xc}}$ that depends on $u$.
In particular, we want to show that the accuracy obtained
with a $u$-dependent functional which does not depend on derivatives of
$\rho$ [i.e., $\varepsilon_{\text{xc}}=\varepsilon_{\text{xc}}(\rho,u)$]
can be similar to the accuracy of GGA functionals
[$\varepsilon_{\text{xc}}=\varepsilon_{\text{xc}}(\rho,\nabla\rho)$].
For this we will consider the functional of Perdew, Burke, and Ernzerhof (PBE)
\cite{PerdewPRL96}, which is one of the standard GGA functionals,
and replace its $\nabla\rho$ dependency by a $u$ dependency.
The analytical form of the PBE functional
$E_{\text{xc}}^{\text{PBE}}=E_{\text{x}}^{\text{PBE}}+E_{\text{c}}^{\text{PBE}}$
is now reviewed in detail.

The non-spin-polarized version of the exchange component of PBE
(the spin-polarized version is trivially calculated \cite{OliverPRA79}) is given by
\begin{equation}
E_{\text{x}}^{\text{PBE}} = \int\varepsilon_{\text{x}}^{\text{LDA}}(r_{s})
F_{\text{x}}^{\text{PBE}}(s)d^{3}r,
\label{ExPBE}
\end{equation}
where $\varepsilon_{\text{x}}^{\text{LDA}}=-\left(9/\left(16\pi\right)\right)
\left(9/\left(4\pi^{2}\right)\right)^{1/3}r_{s}^{-4}$
[$r_{s}=\left(3/\left(4\pi\rho\right)\right)^{1/3}$ is the Wigner-Seitz radius]
is the exchange energy density of the HEG and
\begin{equation}
F_{\text{x}}^{\text{PBE}}(s) = 1 + \kappa - \frac{\kappa}{1 + \frac{\mu}{\kappa}s^{2}}
\label{FxPBE}
\end{equation}
is the beyond-LDA enhancement factor, where $\mu\simeq0.21951$ and $\kappa=0.804$.
PBE correlation is given by
\begin{equation}
E_{\text{c}}^{\text{PBE}} = \int\left[\varepsilon_{\text{c}}^{\text{LDA}}(r_{s},\zeta) +
H^{\text{PBE}}(r_{s},\zeta,t)\right]d^{3}r,
\label{EcPBE}
\end{equation}
where $\varepsilon_{\text{c}}^{\text{LDA}}$ is the correlation energy of the
HEG [$\zeta=\left(\rho_{\uparrow}-\rho_{\downarrow}\right)/\rho$ is the
relative spin polarization],
whose exact analytical form as a function of $\rho$ is unknown, but can
be approximated by a fit of very accurate Monte-Carlo data of the
HEG \cite{CeperleyPRL80,PerdewPRB92a}. The beyond-LDA term in Eq.~(\ref{EcPBE})
is given by
\begin{equation}
H^{\text{PBE}}(r_{s},\zeta,t) = \gamma\phi^{3}
\ln\left(1 + \frac{\beta}{\gamma}t^{2}\frac{1 + At^{2}}{1 + At^{2} + A^{2}t^{4}}\right),
\label{HPBE}
\end{equation}
where
\begin{equation}
t(\mathbf{r})=\left(\frac{3\pi^{2}}{16}\right)^{1/3}\frac{s(\mathbf{r})}
{\sqrt{r_{s}(\mathbf{r})}\phi(\mathbf{r})}
\label{t}
\end{equation}
with $\phi=\left[\left(1+\zeta\right)^{2/3}+\left(1-\zeta\right)^{2/3}\right]/2$,
$\beta=3\mu/\pi^{2}\simeq0.066725$, $\gamma=\left(1-\ln2\right)/\pi^{2}$, and
$A=\left(\beta/\gamma\right)
\left[\exp\left(-\varepsilon_{\text{c}}^{\text{LDA}}/
\left(\gamma\phi^{3}\right)\right)-1\right]^{-1}$.

Our construction of $\varepsilon_{\text{xc}}(\rho,u)$ consists of simply
replacing $s$ by $u-1$ in Eq.~(\ref{FxPBE}) for exchange
and in Eq.~(\ref{t}) for correlation.
This choice is dictated by the requirement that a functional
should recover LDA for the HEG, i.e., when $s=t=0$ for a GGA or $u=1$ for
our approximation. 
Nevertheless, an important point to note is that, while
$s=t=0$ if $\nabla\rho=0$ (since by definition $s$ and $t$ depend
\textit{locally} on $\nabla\rho$), this may not be the
case for $u-1$, since $u$ depends on $\epsilon_{i}$
which in turn depend \textit{nonlocally} (via the KS equations) on $\rho$,
$\nabla\rho$, etc [see Eq.~(\ref{epsiloni})]. Thus, it is only for the HEG
(i.e., $\nabla\rho=0$ $\forall\mathbf{r}$) that one can be sure that $u=1$.
On the other hand, thanks to this nonlocality, $u$ should convey
more or different information than $s$ and $t$.
In this respect, we recall that from the KS equations, the orbital
energies can be expressed as
\begin{eqnarray}
\epsilon_{i} & = &
-\frac{1}{2}\int\psi_{i}^{*}(\mathbf{r})\nabla^{2}\psi_{i}(\mathbf{r})d^{3}r \nonumber \\
& & + \int\left(v_{\text{ext}}(\mathbf{r}) + v_{\text{H}}(\mathbf{r}) +
v_{\text{xc}}(\mathbf{r})\right)\left\vert\psi_{i}(\mathbf{r})\right\vert^{2}d^{3}r,
\label{epsiloni}
\end{eqnarray}
where $v_{\text{ext}}(\mathbf{r})$ is the external potential due to the nuclei,
$v_{\text{H}}(\mathbf{r})=\int\rho(\mathbf{r}')/\left\vert\mathbf{r}-\mathbf{r}'\right\vert d^{3}r'$
is the Hartree potential, and $v_{\text{xc}}(\mathbf{r})$ is the xc potential.
Thus, from Eq.~(\ref{epsiloni}) we can see that the expression for $\epsilon_{i}$
involves nonlocal quantities as $v_{\text{H}}$
(see Ref.~\cite{ConstantinJCP16} for a $v_{\text{H}}$-dependent
exchange functional). Of course, the fact that $s$ and
$u-1$ are not equal also means that the results obtained
after the replacement $s\rightarrow u-1$ will differ from
the original ones. In the following, PBE$u$ refers to the PBE functional with
$s$ substituted by $u-1$. As a technical detail, we mention that depending on
the analytical form of the original GGA, negative values of $u-1$ may lead to
problems. However, this is not the case with PBE since only $s^{2}$ and
$t^{2}$ occur in Eqs.~(\ref{FxPBE}) and (\ref{HPBE}).

In order to know to which extent the gradient-free parameter $u$ can replace
$s$ in a GGA functional or, more generally, can be useful for the future
development of xc functionals, the accuracy of PBE$u$ will be compared to the
accuracy of LDA and GGA functionals. A natural choice for a GGA is PBE
[Eqs.~(\ref{ExPBE})-(\ref{t})], however a certain number of variants of
PBE which differ in the value of the parameters $\mu$, $\kappa$, and
$\beta$ in Eqs.~(\ref{FxPBE}) and (\ref{HPBE}) exist
(see, e.g., Refs.~\cite{ZhangPRL98,PerdewPRL08,HaasPRB10,ConstantinPRL11}).
Among these PBE variants, PBEsol \cite{PerdewPRL08} for which
$\mu=10/81\simeq0.12346$ and $\beta=0.046$ ($\kappa=0.804$ as in PBE), is also
chosen for the comparison with PBE$u$.

\begin{table*}[t]
\caption{\label{table1}The ME, MAE, MRE, and MARE with respect to
experiment \cite{SchimkaJCP11,LejaeghereCRSSMS14} on the testing set of 44
solids for the lattice constant $a_{0}$, bulk modulus $B_{0}$, and
cohesive energy $E_{\text{coh}}$. The units of the ME and MAE are \AA, GPa, and
eV/atom for $a_{0}$, $B_{0}$, and $E_{\text{coh}}$, respectively, and \% for the
MRE and MARE. All results were obtained non-self-consistently using PBE
orbitals/density.}
\begin{ruledtabular}
\begin{tabular}{lcccccccccccc}
\multicolumn{1}{l}{} &
\multicolumn{4}{c}{$a_ {0}$} &
\multicolumn{4}{c}{$B_ {0}$} &
\multicolumn{4}{c}{$E_{\text{coh}}$} \\
\cline{2-5}\cline{6-9}\cline{10-13}
\multicolumn{1}{l}{Functional} &
\multicolumn{1}{c}{ME} &
\multicolumn{1}{c}{MAE} &
\multicolumn{1}{c}{MRE} &
\multicolumn{1}{c}{MARE} &
\multicolumn{1}{c}{ME} &
\multicolumn{1}{c}{MAE} &
\multicolumn{1}{c}{MRE} &
\multicolumn{1}{c}{MARE} &
\multicolumn{1}{c}{ME} &
\multicolumn{1}{c}{MAE} &
\multicolumn{1}{c}{MRE} &
\multicolumn{1}{c}{MARE} \\
\hline
LDA                     &    -0.071 &     0.071 &      -1.5 &       1.5 &      10.1 &      11.6 &       8.1 &       9.5 &      0.78 &      0.78 &      17.5 &      17.5 \\
PBE                     &     0.056 &     0.061 &       1.1 &       1.2 &     -11.1 &      12.2 &      -9.7 &      10.9 &     -0.12 &      0.18 &      -3.7 &       4.8 \\
PBEsol                  &    -0.005 &     0.030 &      -0.1 &       0.6 &       0.8 &       7.8 &      -1.3 &       6.9 &      0.30 &      0.32 &       6.4 &       7.0 \\
PBE$u$(PBE)             &     0.018 &     0.048 &       0.3 &       1.1 &       2.3 &      10.3 &      -4.1 &      11.4 &     -0.47 &      0.65 &      -9.3 &      13.2 \\
PBE$u$(PBEsol)          &    -0.036 &     0.040 &      -0.8 &       0.9 &       7.8 &      11.7 &       2.1 &       8.7 &      0.11 &      0.45 &       3.3 &       9.3 \\
PBE$u$(0.10,0.02)       &    -0.024 &     0.030 &      -0.6 &       0.7 &       4.5 &       8.5 &       0.5 &       7.8 &      0.06 &      0.35 &       1.3 &       7.1 \\
\end{tabular}
\end{ruledtabular}
\end{table*}

\begin{figure}
\includegraphics[width=\columnwidth]{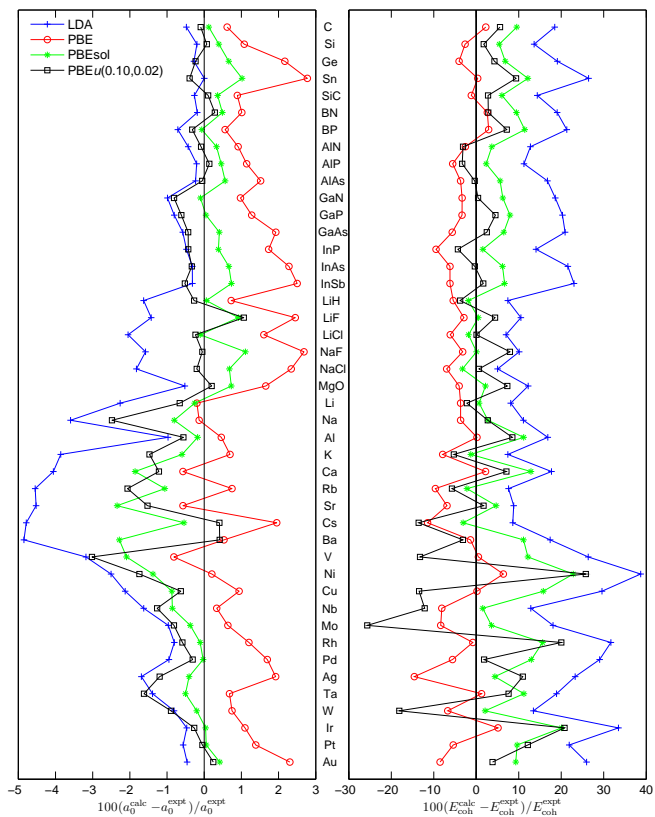}
\caption{\label{fig_solids}Relative error (in \%) with respect to
experiment \cite{SchimkaJCP11,LejaeghereCRSSMS14} in the calculated lattice
constant (left panel) and cohesive energy (right panel) for 44 solids.}
\end{figure}

The xc functionals will be compared for their accuracy on the equilibrium
lattice constant $a_{0}$, bulk modulus $B_{0}$, and cohesive energy
$E_{\text{coh}}$ of solids. The test set is the one
that we used in our previous work \cite{TranJCP16} and consists of
44 cubic solids of various types ($sp$ semiconductors,
ionic insulators, and metals).
The calculations were done with the WIEN2k code \cite{WIEN2k},
which is based on the LAPW method \cite{Singh}.
The results are shown in Tables~S1-S3 and
Figs.~S1-S6 of Ref.~\cite{SM_new_variable}, while
Table~\ref{table1} shows the mean error (ME),
mean absolute error (MAE), mean relative error (MRE), and mean absolute
relative error (MARE) with respect to experiment.
All results were obtained non-self-consistently by using the density $\rho$
and orbitals $\psi_{i}$ (and $\epsilon_{i}$) generated
by a PBE calculation. We checked that using the density, etc.
from LDA leads to negligible changes in the results.

It is known that LDA strongly underestimates (overestimates) the lattice constant
(cohesive energy), and from Fig.~S1 we can see that it is the case for all solids.
The MAE obtained with LDA amount to $0.071$~\AA~and
$0.78$~eV/atom for $a_{0}$ and $E_{\text{coh}}$, respectively, and are the largest
among all tested functionals. The bulk modulus is overestimated
for the vast majority of solids and the MAE is 11.6~GPa.
On average, PBE is only slightly more accurate than LDA for the lattice constant
($\text{MAE}=0.061$~\AA), while
the reverse is observed for the bulk modulus ($\text{MAE}=12.2$~GPa).
From Fig.~S2 and the ME, we can see that the tendency of PBE is to
overestimate $a_{0}$ (and therefore to underestimate $B_{0}$).
For the cohesive energy, PBE is much more accurate than LDA since
the MAE is four times smaller (0.18~eV/atom).
PBEsol, which was proposed as a more accurate GGA for the lattice constant of solids
\cite{PerdewPRL08}, leads to MAE for $a_{0}$ (0.030~\AA) and $B_{0}$ (7.8~GPa)
that are clearly smaller than for PBE. However, PBEsol is less accurate than PBE for
$E_{\text{coh}}$ ($\text{MAE}=0.32$~eV/atom).

Turning to the PBE$u$ functional, Table~\ref{table1} shows the results
obtained with three variants of PBE$u$, which differ in the values of
$\mu$ and $\beta$ in Eqs.~(\ref{FxPBE}) and (\ref{HPBE}),
respectively ($\kappa=0.804$ for all functionals). In PBE$u$(PBE) and
PBE$u$(PBEsol), the PBE and PBEsol parameters mentioned above are used,
while PBE$u$(0.10,0.02) is a reparametrization with $\mu=0.10$ and
$\beta=0.02$. For $a_{0}$, PBE$u$(PBE) and PBE$u$(PBEsol) lead to
values of 0.048 and 0.040~\AA~for the MAE, such that their overall accuracy is
somewhere in between PBE and PBEsol. The MAE for $B_{0}$ obtained with
these two functionals (10.3 and 11.7~GPa) are quite similar to the values obtained
with LDA and PBE, but larger than for PBEsol. With MAE of 0.65 and 0.45~eV/atom
for $E_{\text{coh}}$, PBE$u$(PBE) and PBE$u$(PBEsol) are superior to LDA,
but clearly inferior to PBE which is the most accurate functional tested in
this work for $E_{\text{coh}}$. Thus, by considering overall the MA(R)E
for the three properties, PBE$u$(PBE) and PBE$u$(PBEsol), which are constructed by
just replacing $s$ by $u-1$, improve over LDA the same way as PBE and PBEsol do.

\begin{figure*}
\includegraphics[scale=0.6]{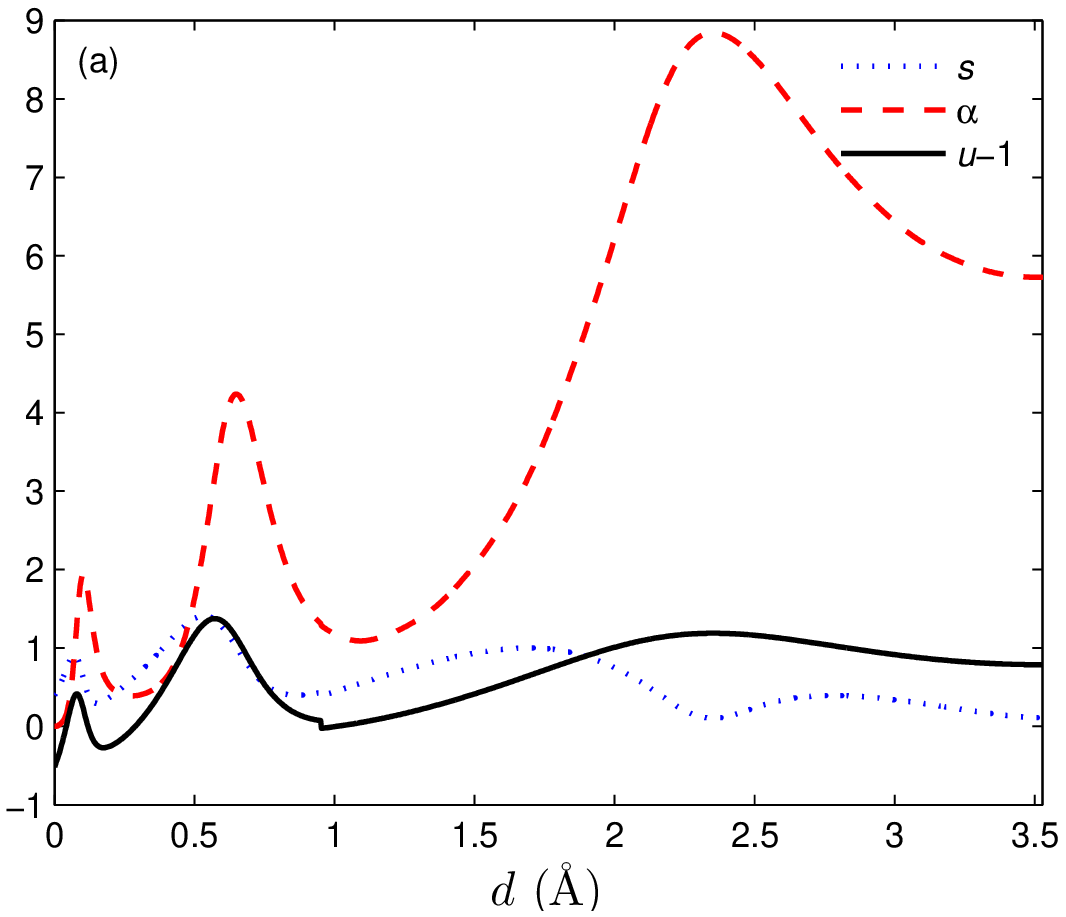}
\includegraphics[scale=0.6]{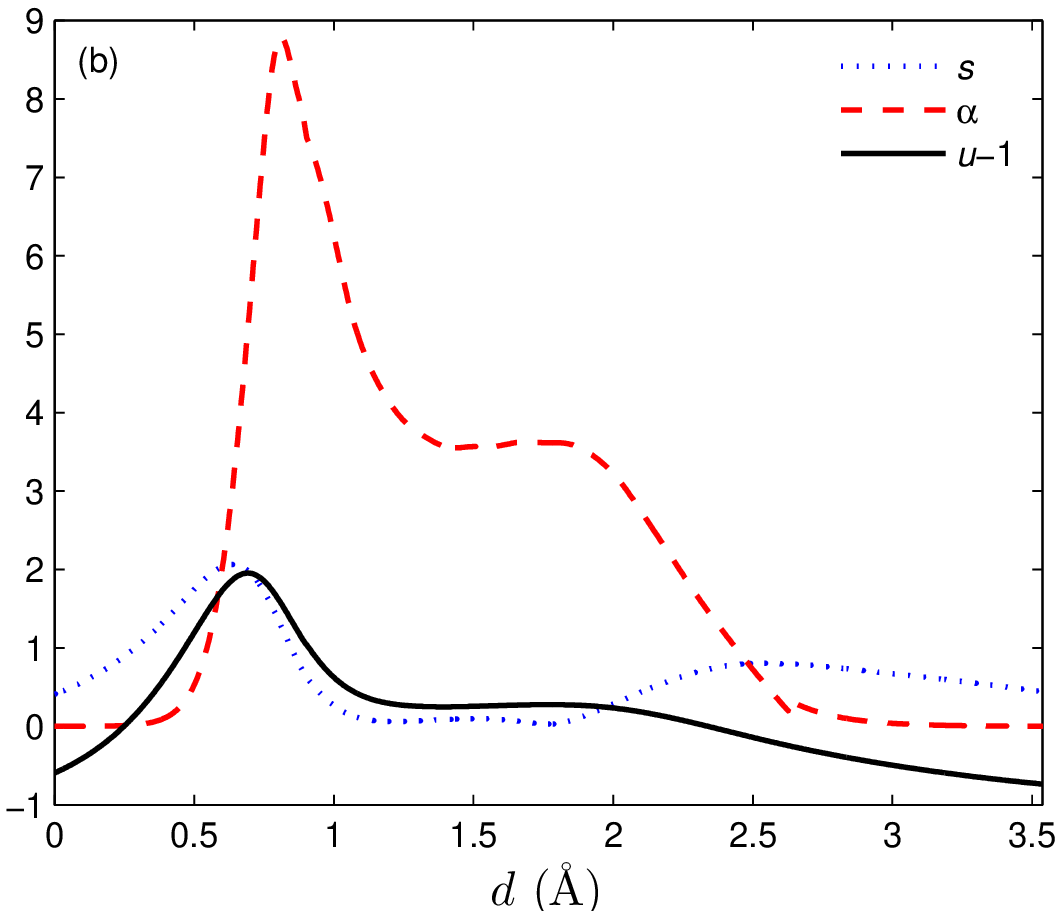}
\includegraphics[scale=0.6]{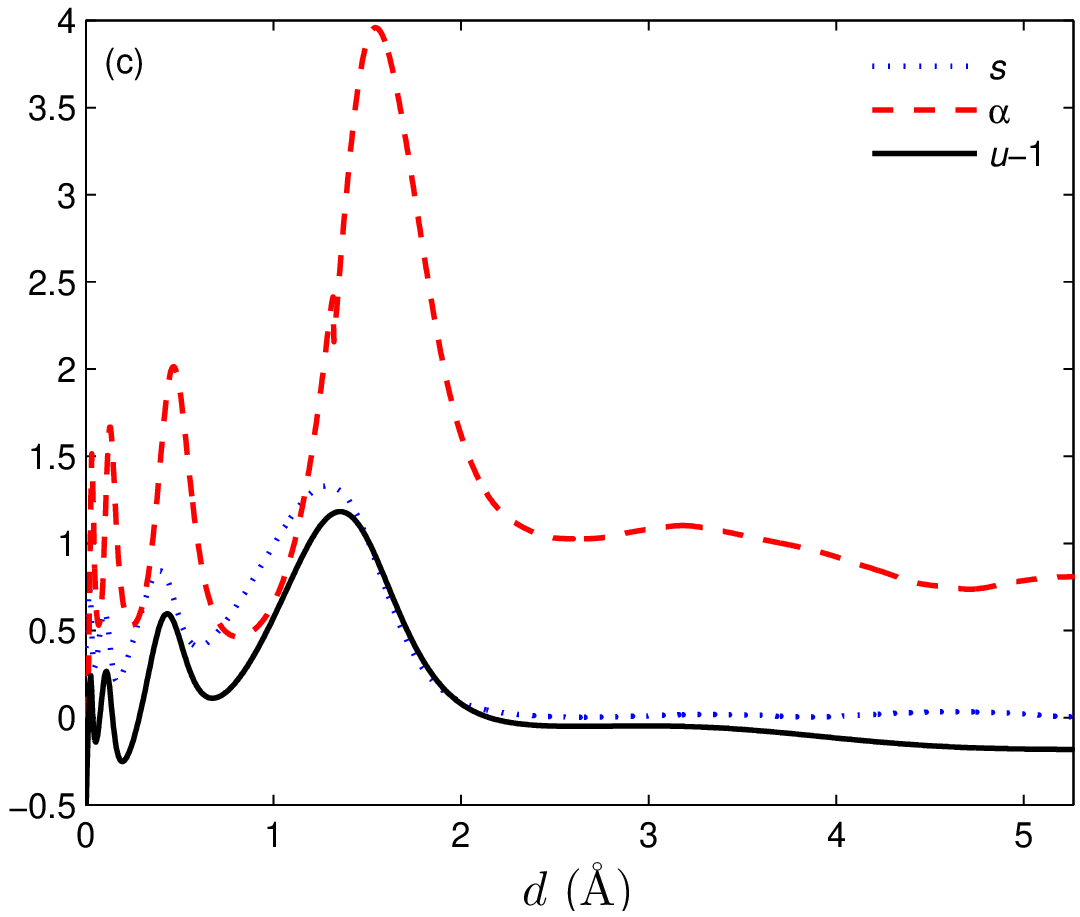}
\includegraphics[scale=0.6]{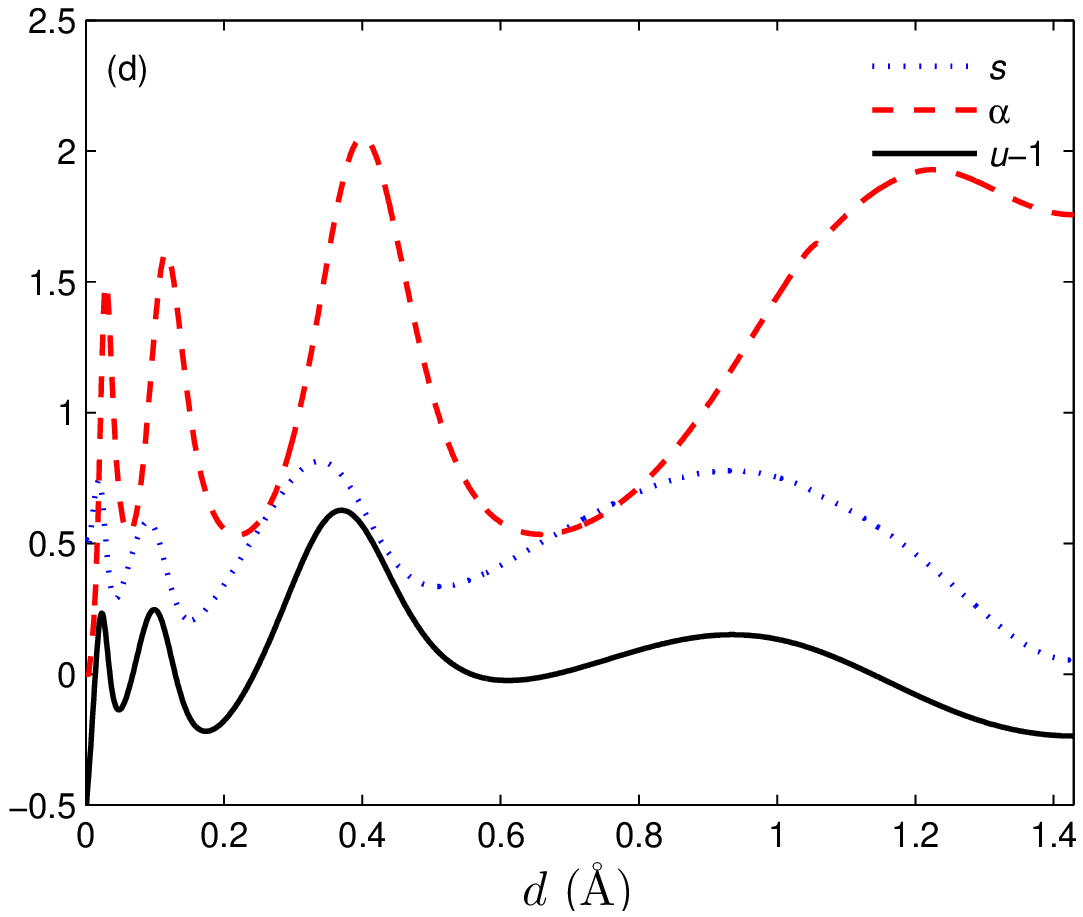}
\caption{\label{fig_fun}$s$ [Eq.~(\ref{s})], $\alpha$
[Eq.~(\ref{alpha})], and $u-1$ [Eq.~(\ref{u})] in
(a) Si from the atom at $(1/8,1/8,1/8)$ to $(1/2,1/2,1/2)$,
(b) LiH from the Li atom at $(0,0,0)$ to the H atom at $(1/2,1/2,1/2)$,
(c) Sr from the atom at $(0,0,0)$ to $(1/2,1/2,1/2)$, and
(d) Nb from the atom at $(0,0,0)$ to $(1/4,1/4,1/4)$.}
\end{figure*}

\begin{figure}
\includegraphics[width=\columnwidth]{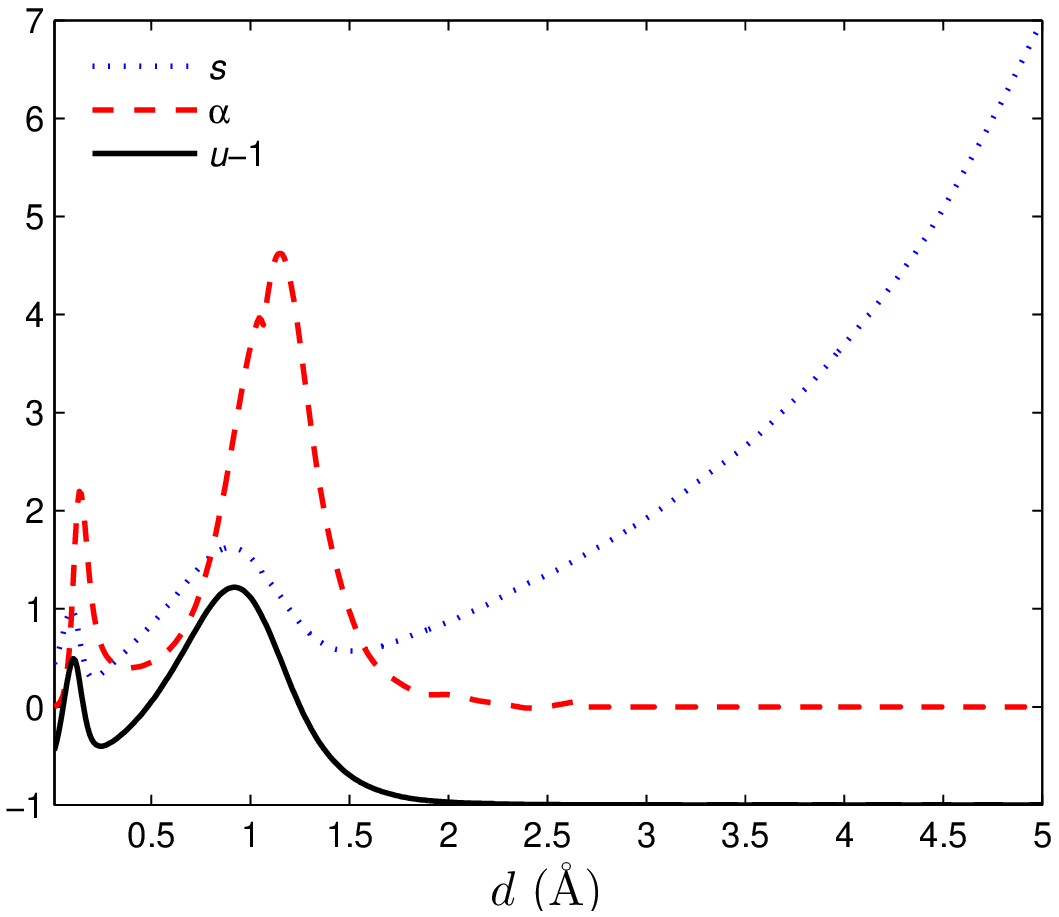}
\caption{\label{fig_fun2}Majority-spin component of $s$
[Eq.~(\ref{s})], $\alpha$ [Eq.~(\ref{alpha})], and $u-1$ [Eq.~(\ref{u})]
in an isolated Na atom.}
\end{figure}

However, the results with PBE$u$ can be improved by tuning
$\mu$ and $\beta$ and a combination, $\mu=0.10$ and $\beta=0.02$, leads to errors which
are reduced. From Table~\ref{table1}, we can see that
the MAE is 0.030~\AA, 8.5~GPa, and 0.35~eV/atom for $a_{0}$, $B_{0}$, and
$E_{\text{coh}}$, respectively. Thus, the accuracy achieved by
PBE$u$(0.10,0.02) is overall similar to PBEsol accuracy.

Looking in more detail at the results, Fig.~\ref{fig_solids} shows
the results for the lattice constant and cohesive energy.
Several observations can be made.
Starting with the $sp$ semiconductors (i.e., from C to InSb), we can see that
the values of $a_{0}$ obtained with LDA, which are quite accurate (the errors are
similar to PBEsol and much smaller than PBE), are followed closely by
PBE$u$(0.10,0.02) results. However, PBE$u$(0.10,0.02) improves significantly
over LDA for the cohesive energy of the $sp$ semiconductors and shows similar
accuracy as PBE. Concerning the ionic solids (i.e., from LiH to MgO) the errors for
$a_{0}$ are overall the smallest with PBE$u$(0.10,0.02), while for
$E_{\text{coh}}$ the PBE$u$(0.10,0.02) errors are on average of similar magnitude
as with PBE, but with opposite sign. For Al and the alkali and alkaline earth metals,
the magnitude of the errors with PBE$u$(0.10,0.02) and PBEsol are rather similar
for both $a_{0}$ and $E_{\text{coh}}$. The most visible exceptions are Na and Ba
for $a_{0}$. The PBE$u$(0.10,0.02) lattice constants for the transition metals
lie in between the LDA and PBEsol results. PBE is the most accurate method for the
$3d$ metals, while PBEsol is recommended for the $5d$ metals.
Regarding the cohesive energy of the transition metals, PBE is the most accurate
method, while PBEsol is overall somewhat less accurate and systematically
overestimates the values. However, for the cohesive energy
PBE$u$(0.10,0.02) leads for a few cases to large deviations from experiment.
For V, Cu, Nb, Mo, and W, $E_{\text{coh}}$ is clearly underestimated, while
large overestimations similar to PBEsol are obtained for Ni, Rh, and Ir.

Figure~\ref{fig_fun} shows plots of $s$, $\alpha$, and $u-1$ for
selected solids. We can see that the positions of the peaks in
$s$ and $u$ coincide well despite $s$ depends
on $\nabla\rho$, while $u$ does not. These similar features
explain why substituting $s$ by $u-1$ in a GGA leads to a
functional that can also be much more accurate than LDA.
Nevertheless, differences between $s$ and $u-1$
can also be observed. For instance, $s-u$ is not constant,
which is more clearly visible in LiH
[Fig.~\ref{fig_fun}(b)]. Also, in Si [Fig.~\ref{fig_fun}(a)]
$s$ and $u$ show opposite curvatures at $d\sim2.3$~\AA~and
$u-1$ is clearly larger than $s$ in the interstitial.
Actually, such differences indicate that $u$ should be
considered as a complementary variable to $s$ and $\alpha$ for
functionals development.
The peaks of $\alpha$ are at slightly different positions, which are
shifted far away from the nucleus compared to $s$ and $u-1$.
However, far from nuclei, where the density tail is,
$s$ and $u-1$ differ drastically. Figure~\ref{fig_fun2}
shows their majority-spin component in an isolated Na atom,
where we can see that starting from $d\sim1.5$~\AA, $s$ increases
(with $\lim_{d\to\infty}s=\infty$), while $u$ goes to zero.

In summary, we have shown that $u$, as defined by Eq.~(\ref{u}), can be used
as a variable in xc functionals to improve the results over the LDA functional.
Taking PBE as an example,
we have shown that the accuracy of the $u$-dependent functional PBE$u$ can be made
as accurate as the standard GGAs like PBE or PBEsol which depend on $s$.
What is remarkable is that $u$ does not depend explicitly on any derivative of the
density $\rho$ but is able to detect inhomogeneities in $\rho$
pretty much the same way as $s$ does. Furthermore,
since $u$ is a more nonlocal quantity than $s$, it should carry more information
and therefore be a useful complement to $s$ and $\alpha$ for the future development
of more accurate xc functionals. To finish we mention that
the $\epsilon_{\text{H}}$ dependency may lead to a simple way of
calculating the derivative
discontinuity (relevant for the band gap) with the total energy \cite{YangJCP12}
in the same spirit as done with the GLLB potential \cite{KuismaPRB10}.
Concerning the functional derivative, the use of
the chain rule, either $\delta\epsilon_{j}/\delta\rho$ (in the KS scheme \cite{EngelJCC99}) or
$\delta\epsilon_{j}/\delta\psi_{i}^{*}$ (in the generalized KS scheme \cite{SeidlPRB96}),
would be needed. Either way, its calculation is less straightforward than for
semilocal functionals and should require the calculation of the
response function. However, how complicated to implement or expensive
such a method would be is at present unclear.

\begin{acknowledgments}

This work was supported by the project F41 (SFB ViCoM) of the Austrian Science
Fund (FWF).

\end{acknowledgments}

\bibliography{references}

\end{document}